\documentclass[conference,12pt]{IEEEtran}

\IEEEoverridecommandlockouts
\usepackage{cite}
\usepackage{amsmath,amssymb,amsfonts}
\usepackage{algorithmic}
\usepackage{graphicx}
\usepackage{textcomp}
\usepackage{xcolor}

\begin{document}
\title{Intrusion Detection using Sequential Hybrid Model}

\author{\IEEEauthorblockN{Aditya Pandey}
\IEEEauthorblockA{\textit{Department of Computer Science} \\
\textit{PES University}\\
Bengaluru, India - 560085\\}
\and
\IEEEauthorblockN{Abhishek Sinha}
\IEEEauthorblockA{\textit{Department of Computer Science} \\
\textit{PES University}\\
Bengaluru, India - 560085\\}
\and
\IEEEauthorblockN{Aishwarya PS}
\IEEEauthorblockA{\textit{Department of Computer Science} \\
\textit{PES University}\\
Bengaluru, India - 560085\\}
}
\date{\today}
\maketitle

\begin{abstract}
\\A large amount of work has been done on the KDD 99 dataset, most of which includes the use of a hybrid anomaly and misuse detection model done in parallel with each other. In order to further classify the intrusions, our approach to network intrusion detection includes use of two different anomaly detection models followed by misuse detection applied on the combined output obtained from the previous step. \\The end goal of this is to verify the anomalies detected by the anomaly detection algorithm and clarify whether they are actually intrusions or random outliers from the trained normal (and thus to try and reduce the number of false positives). We aim to detect a pattern in this novel intrusion technique  itself, and not the handling of such intrusions. The intrusions were detected to a very high degree of accuracy. 
\end{abstract}

\section{Introduction}
\subsection{Context}

A network intrusion could be any unauthorised activity on a computer network. Virus attacks, unauthorised access, theft of information and denial-of-service attacks were the greatest contributors to computer crime. Detecting an intrusion depends on the defenders having a clear understanding of how attacks work. Detecting an intrusion is the first step to create any sort of counteractive security measure. Hence, it is very important to accurately determine whether a connection is an intrusion or not.

\subsection{Categories}

There are four broad categories of intrusions in a network of systems:
\begin{itemize}
\item DOS: \\In computing, a denial-of-service attack is a cyber-attack in which the perpetrator seeks to make a machine or network resource unavailable to its intended users by temporarily or indefinitely disrupting services of a host connected to the Internet. E.g. back, land, neptune.
\item U2R: \\These attacks are exploitations in which the hacker starts off on the system with a normal user account and attempts to abuse vulnerabilities in the system in order to gain super user privileges. E.g. perl, xterm.
\item R2L: \\Remote to local is an unauthorised access from a remote machine, by maybe guessing the password of the local system and accessing files within the local system. E.g. ftp\_write, guess\_passwd, imap.
\item Probe: \\Probing is an attack in which the hacker scans a machine or a networking device in order to determine weaknesses or vulnerabilities that may later be exploited so as to compromise the system. This technique is commonly used in data mining. E.g. saint, portsweep, mscan, nmap etc.
\end{itemize}

\subsection{Importance of Intrusion Detection}
Intrusion Detection is important for both Military as well as commercial sectors for the sake of their Information Security, which is the most important topic of research for the future networks. It is critical to maintain a high level of security to ensure safe and trusted communication of information between various organisations. \\Intrusion Detection System is a new safeguard technology for system security after traditional technologies, such as firewall, message encryption and so on. An intrusion detection system (IDS) is a device or software application that monitors network system activities for malicious activities or policy violations and produces reports to a Management Station.

\section{Previous Work}
The two approaches to an Intrusion Detection System are Misuse Detection and Anomaly Detection. A key advantage of misuse detection techniques is its high degree of accuracy in detecting known attacks and their variations. Their obvious drawback is the inability to detect attacks whose instances have not yet been observed. Anomaly detection schemes on the other hand, suffer from a high rate of false alarms. This occurs primarily because previously unseen (yet legitimate) system behaviours are also recognized as anomalies, and are hence flagged as potential intrusions. Researchers have used various different algorithms to solve this specific problem. \\
A data mining algorithm such as Random Forest can be applied for misuse, anomaly and hybrid network-based intrusion detection systems. ~\cite{b7} uses a hybrid detection technique where they employ misuse detection followed by anomaly detection. This approach however has multiple limitations:
\begin{itemize}
\item Intrusions need to be much lesser than the normal data. Outlier detection will only work when majority of the data is normal
\item A novel intrusion producing a large number of connections that are not filtered out by the mis- use detection could decrease the performance of the anomaly detection or even the hybrid system as a whole.
\item Some of the intrusions with a high degree of similarity cannot be detected by this anomaly detection approach.
\end{itemize}
The paper 'Artificial Neural Networks for Misuse Detection' ~\cite{b8} talks about the advantages of using an artificial neural network approach over a rule based expert system approach. It also tells us about the various approaches with which these neural networks are applied to get a high accuracy.
A review paper on misuse detections ~\cite{b9} sheds light on the most common techniques to implement misuse detection. These are:
\begin{itemize}
\item Expert Systems, which code knowledge about attacks as 'if-then' implication rules.
\item Model Based Reasoning Systems, which combine models of misuse with evidential reasoning to support conclusions about the occurrence of a misuse.
\item State Transition Analysis, which represents attacks as a sequence of state transitions of the monitored system.
\item Key Stroke Monitoring, which uses user key strokes to determine the occurrence of an attack. 
\end{itemize}
~\cite{b9} introduces a pattern matching model along with an Artificial Neural Network model for their implementation of a Misuse Detection System. Further ~\cite{b9} performed Intrusion Detection using Data Mining techniques along with fuzzy logic and basic genetic algorithms.
~\cite{b1},~\cite{b2},~\cite{b3},~\cite{b4},~\cite{b5} are varied approaches but their accuracy is not as good as the approach in ~\cite{b7}. The approach specified in ~\cite{b6} has a good accuracy and also supports unsupervised learning algorithms but it requires a very complicated state machine specification model.

\section{Problem Statement}
Here, our goal is to detect network intrusions and to create a predictive model that is able to differentiate between 'good' or 'bad' connections (intrusions) and classify those intrusions into known categories. To that end, we used the kddCup dataset from 1999 which was created by simulating various intrusions over a network over the course of several weeks in a military environment. \\The dataset itself consists of 42 attributes that are used in the various models according to relevance and importance. It contains approximately 1,500,000 data-points. For the purposes of this project, we use a 10\% representative dataset, since our machines did not have the processing power to handle the larger dataset. The dataset as is, is unbalanced across the various result categories, and hence, we balance it by up-sampling and down-sampling. The dataset is also cleaned. Certain categorical variables like the 'protocol\_type' column are one-hot encoded. We found out that the 'flag' and 'services' attributes are not of much value as they are nominal variables and hence, they are dropped. 
\begin{table}[ht]
\caption{Number of Data Points} 
\centering 
\begin{tabular}{c c c c c c}
\hline
Label: & dos & normal & probe & rtl & u2r \\
\hline
Before Sampling: & 54572 & 87832 & 2131 & 999 & 52  \\ 
After Sampling: & 27285 & 39524 & 2131 & 999 & 86 \\
\hline 
\end{tabular}
\end{table}

\section{Approach}
Due to the various drawbacks of the individual anomaly detection models as well as the individual misuse detection models, we use a combined approach i.e. a hybrid of the two. We combine the models serially such that the anomaly detection is followed by the misuse detection. This approach provides us with multiple advantages:
\begin{itemize}
\item The misuse detection acts as a verification model where it verifies whether an anomaly is actually an intrusion or not. This helps us reduce the false positives coming from the anomaly detection (Primary drawback)
\item The misuse detection also helps us classify the intrusions detected into various categories based on the intrusions' 'signature' (centroid of sample data from the training dataset)
\end{itemize}
\begin{figure}[!htb]
  \includegraphics[width=\linewidth]{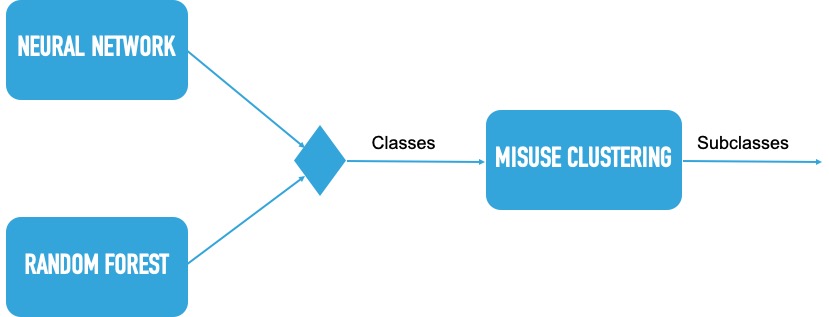}
  \caption{Block Diagram of Approach}
  \label{fig:BlkDiag}
\end{figure}
Figure \ref{fig:BlkDiag} shows the approach for the Intrusion Detection System.

\subsection{Modelling}
For the anomaly detection, we use two models:
\begin{itemize}
\item Random Forest Model
\item Neural Network Model
\end{itemize}
After running both these classification models simultaneously, we take the combined output of the anomalies detected by either of the two. Doing so helps us reduce the rate of false negatives. The connections which are identified as anomalies using either of the two are then passed on to the misuse detection model. A false positive from the anomaly detection model classified as 'normal' in the misuse detection model reduces the number of false positives.\\
For Misuse detection, we use:
\begin{itemize}
\item KMeans based Clustering Model.
\end{itemize}
Therefore, the use of the misuse detection model on the anomalies helps trim down the number of false positives.

\section{Components of the Intrusion Detection System}

Our Intrusion Detection System consists of 3 components which have been combined to create one complete robust system.\\ \\
The components of the Anomaly Detection System are - 
\subsection{Neural Network:} 
A neural network is a system of hardware and/or software patterned after the operations of neurons in the human brain.\\For the neural network, we created a sequential neural network model with two hidden layers. The neural network consists of 41 input nodes and 5 output nodes. The 41 input nodes are the numerical attributes which are obtained after pre-processing and the 5 output nodes are used to classify the data point into one of the 5 classes of connections that we have (u2r, rtl, normal, dos, probe).\\The model is validated by a K fold cross validation with k as 2. An average accuracy of 99.57\% was obtained. In an attempt to reduce processing time without much drop in accuracy, we reduced the number of splits in the k fold cross validation as well as used a lower number of hidden layers in the network with minimal loss of accuracy. 

\begin{figure}[!htb]
  \includegraphics[width=\linewidth]{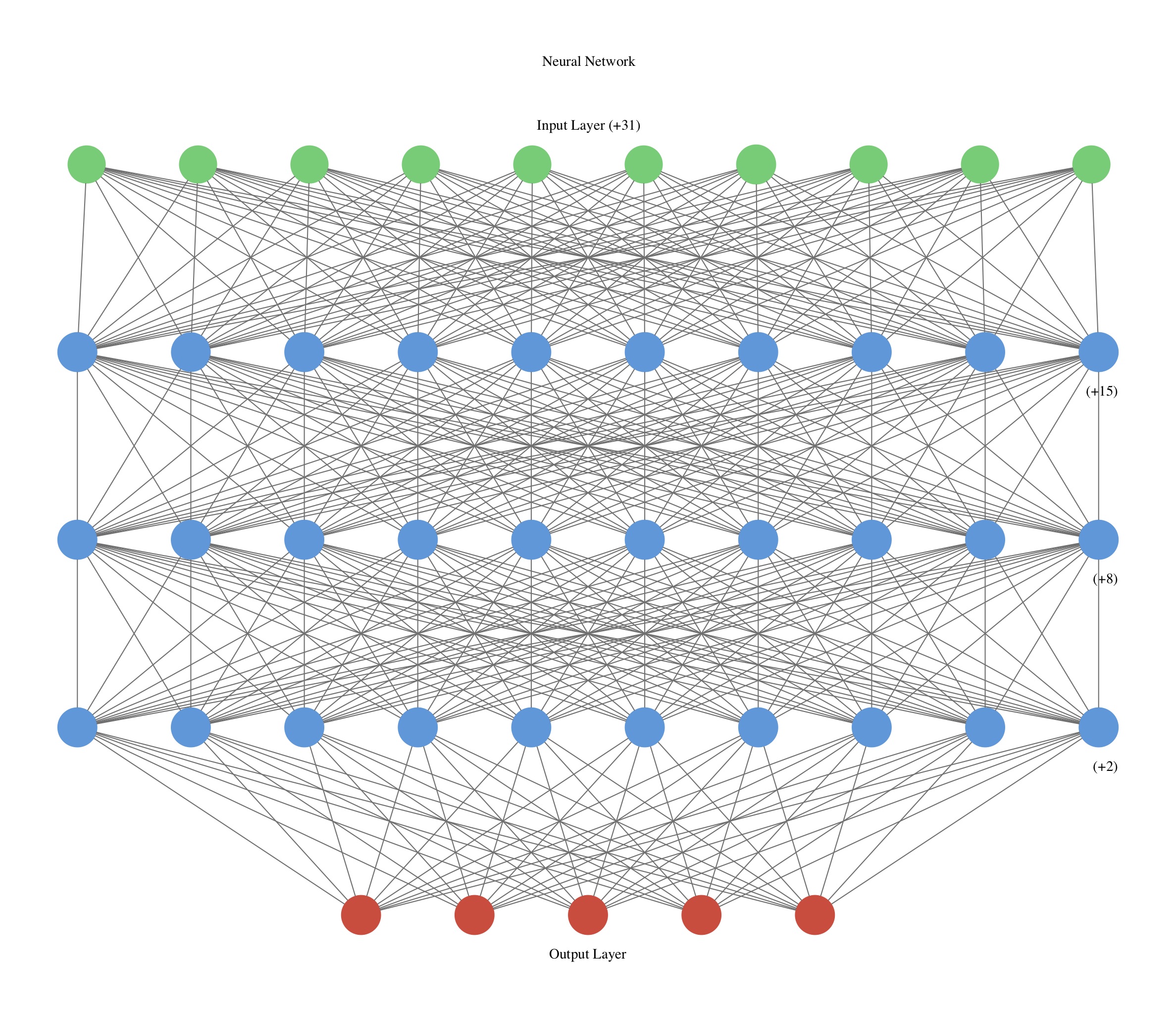}
  \caption{Neural Network Model}
  \label{fig:Nnet}
\end{figure}

Figure \ref{fig:Nnet} shows the number of nodes in each layer of the Neural Network.

\subsection{Random Forest:} A random forest is essentially a multitude of decision trees. The output obtained from a random forest model is a combination of the outputs obtained from all the decision trees.\\In our case, we use the mode of all the decision tree outputs as the final output from the model. Based on trial and error, the number of decision trees in our random forest was taken as 100. We use this Random Forest Classifier to classify the connection among the 5 main classes. \\Based on the importance value of variables, all the non-essential variables are dropped before the model is retrained. This helps in removing all the redundant and useless attributes in the model. Using the random forest model, an average accuracy of 99.78\% is obtained.\\In order to reduce the processing time, the number of decision trees was reduced from an initial value of 1000 to 100 without resulting in any significant loss in accuracy of the model.

\begin{figure}[!htb]
  \includegraphics[width=\linewidth]{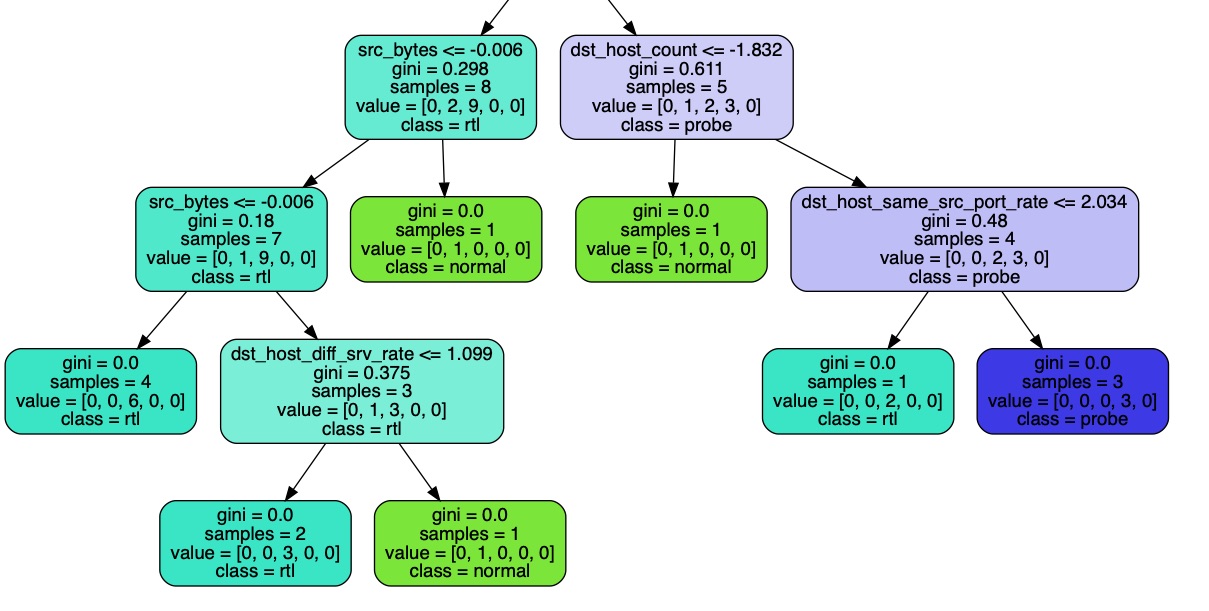}
  \caption{Part of the RF Model}
  \label{fig:rf}
\end{figure}

Figure \ref{fig:rf} shows a part of the large Random Forest Model created

\subsection{Misuse Clustering:} We used a K-Means based clustering algorithm as the misuse detection model. This clustering algorithm is used to further classify the five classes of connections into 24 classes. During training, clusters are made for each known class and their centroids are stored. When the model receives a data point during testing, the minimum distance to any of the centroids is calculated to assign the point to the relevant class.\\ 
If a connection arrives during testing that is falsely tagged as an attack, then the cluster it is assigned to should theoretically be 'normal', thereby reducing false positives. 
\subsection{Combination of Models:} The output from both the models in the anomaly detection stage is verified with each other and the connections labelled as some kind of attack or connections not having the same labels are passed on to the misuse clustering model which will further classify the attack into its subclass. Another version of our final model is to use the Misuse clustering model to reduce the number of false positives, thereby increasing the precision and accuracy.\\

\section{Results}

\begin{table}[ht]
\caption{Neural Network:} 
\centering 
\begin{tabular}{c c c c c c}
\hline
Label: & normal & rtl & dos & u2r & probe \\
\hline
Precision: & 98.391 & 97.560 & 99.955 & 79.710 & 97.468  \\ 
Recall: & 99.817 & 80.080 & 99.146 & 63.953 & 90.333 \\
Accuracy: & 98.976& 99.687 & 99.650 & 99.935 & 99.634 \\
\hline 
\end{tabular}
\end{table}

\begin{table}[ht]
\caption{Random Forest:} 
\centering 
\begin{tabular}{c c c c c c}
\hline
Label: & normal & rtl & dos & u2r & probe \\
\hline
Precision: & 99.820 & 99.382 & 99.985 & 87.500 & 99.669  \\ 
Recall: & 99.949 & 96.596 & 99.978 & 81.395 & 98.967 \\
Accuracy: & 99.870& 99.942 & 99.985 & 99.962 & 99.958 \\
\hline 
\end{tabular}
\end{table}

Note: All values are with respective to individual classes vs the entire dataset.\\

\begin{table}[ht]
\caption{Misuse:} 
\centering 
\begin{tabular}{c c c}
\hline
Type of Classification: & 5 Class & 24 Class\\
Accuracy: & 99.651& 91.31 \\
\hline 
\end{tabular}
\end{table}

\begin{figure}[!htb]
  \includegraphics[width=\linewidth]{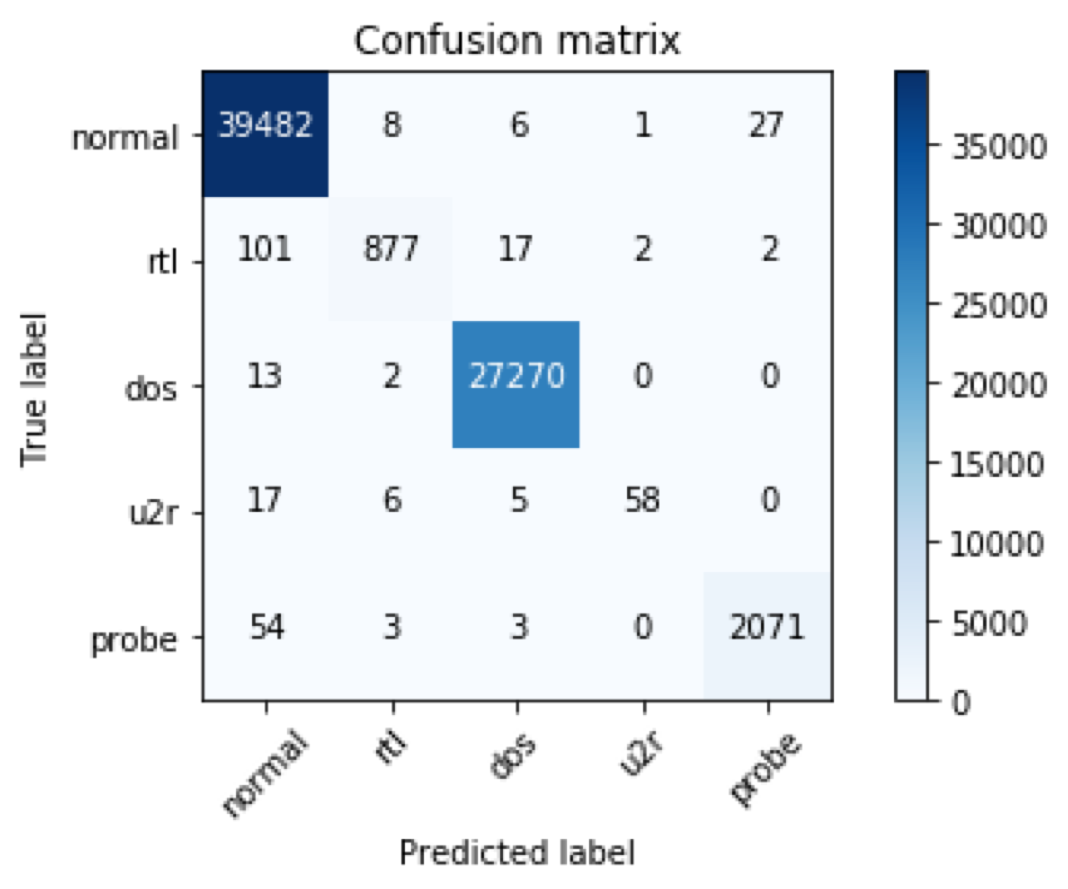}
  \caption{Confusion Matrix for 5 classes}
  \label{fig:finalcm}
\end{figure}

Figure \ref{fig:finalcm} is the confusion matrix of the 5-class misuse classification (the 24-class classification is difficult to visualise).

\subsection{Discussion of Results}
Since we are using a synthetic dataset, we obtain accuracies that are quite high. However, the precision and recall for u2r in specific, in both the models is very low compared to the rest of our numbers. This can be attributed to the fact that the number of connections causing u2r attacks itself is very low, taking up only around 70 of the 70,000 connections. As a result, the models have not been trained enough to be able to detect these attacks. Hence a good proportion of the u2r attacks have been misclassified. This problem can be rectified if a more balanced dataset is provided.\\
From the tabular columns we can see that most of the attacks have not only been detected but also correctly classified in the anomaly detection model. In an attempt to further classify the attacks into its subcategories, we lose a bit of accuracy but it is still at an acceptable 91.3\%. A further classification is necessary from the perspective of dealing with these intrusions. For example, A and B might both belong to 'dos' but they might have different security requirements to stop these attacks. We can also use the misuse clustering to cut down on false positives and that will only make our accuracy and precision better, if in case the drop in accuracy is not preferable.\\
From the confusion matrix, we can also see that the number of 'normal' and 'dos' connections are significantly higher than the number of all the other connections even after down-sampling these categories. Therefore, in order to achieve a balance that is acceptable, we down-sampled these 2 categories and up-sampled the 'u2r' category. \\What can also be seen is that the number of misclassifications is fairly less. The only issue which can be seen is that a fraction of the misclassification is classifying one of the kinds of attacks as normal connections. The most probable reason for this is due to the fact that the number of 'normal' connections make up more than 55\% of the total dataset. However, considering the number of false negatives versus the total number of connections, this error is almost insignificant. 

\section{Conclusion}
Even though technology has advanced leaps and bounds over the past few decades, Intrusion detection continues to be an active research field. This paper outlines our approach to solving this problem by using a combination of Anomaly and Misuse Detection models. The techniques used to perform the same have been explained and illustrated. We believe that this approach is quite successful in classifying connections into their respective categories. 

\section{Future Work}
The current approach is able to spot and classify new intrusions as such. However, the number of false positives for this are high. Furthermore, the connections detected as new intrusions could be further categorised into subclasses based on similarity(clustering) and dealt with accordingly. Different approaches could also be tried in order to obtain more accurate signatures for the different categories of intrusions which would improve the accuracy of the misuse detection model.

\section{Contributions}
This project was done by Abhishek Sinha, Aditya Pandey and Aishwarya PS under the guidance of Dr. Gowri Srinivasa.\\
Parts of the pre-processing of the data was carried out by all of the team members. Abhishek Sinha was responsible for one hot encoding and dropping of certain columns, Aditya Pandey for the up-sampling, down-sampling, removal of duplicate data points and cleaning of the data, and Aishwarya PS adding a column for labelling of various intrusion classes as their parent intrusion domain and for N-Fold cross validation.\\
From here onwards, Aishwarya PS carried out the Random Forest Model, Aditya Pandey carried out the Neural Network Model and Abhishek Sinha carried out the Misuse Detection Model before the approaches were combined.

\end{document}